\documentstyle[aps,preprint]{revtex}

\def \grd {\mbox {\boldmath $\nabla$} }

\def \bsigma {\mbox {\boldmath $\sigma$} }

\begin{document}
\draft
\preprint{}
\title  {Hydrodynamic fluctuations in the Kolmogorov flow: \\ Linear regime}
\author{ I. Bena$^{a}$,  M. Malek  Mansour$^{b}$  and F.  Baras$^{b}$}
\address{(a)  Limburgs Universitair Centrum\\ B-3590 Diepenbeek, Belgium \\
(b) Centre for
Nonlinear Phenomena and Complex Systems\\ Universit\'e Libre de Bruxelles,
Campus Plaine,
C.P. 231 \\ B-1050 Brussels, Belgium }
\date{\today}
\maketitle

\begin{abstract}
The Landau-Lifshitz fluctuating hydrodynamics is used to study the
statistical properties of
the  linearized Kolmogorov flow.  The relative simplicity of this flow
allows a detailed
analysis of the fluctuation spectrum from near equilibrium regime up to the
vicinity of the
first convective instability threshold.  It is shown that in the long
time limit the flow behaves as an incompressible fluid, regardless of the
value of the
Reynolds number.  This is not the case for the short time behavior where
the incompressibility assumption leads in general to  a wrong  form of the
static correlation
functions, except near the instability threshold.  The theoretical
predictions are
confirmed by numerical simulations of the full nonlinear fluctuating
hydrodynamic equations.

\end{abstract}
\


\section{Introduction}
\label{sec:intro}

A common theoretical approach for the study of fluctuations is the
Landau-Lifshitz fluctuating
hydrodynamics \cite{Landau} mainly because of its relative simplicity as
compared to more
fundamental approaches \cite{Others1,Dufty}.  Fluctuating hydrodynamics is
a stochastic
formulation of standard fluid mechanics. Spontaneous fluctuations of
hydrodynamic variables
are introduced into the transport equations by adding random components to
the dissipative
part of the pressure and heat fluxes. Since these fluxes are not conserved
quantities, the
correlations of the random terms are expected to be short-ranged and
short-lived, so that on a
hydrodynamic scale they are assumed to be Dirac-delta correlated.  Their
strengths are then
chosen to yield the correct equilibrium thermodynamic fluctuations as
derived from the Gibbs
distribution.

Fluctuating hydrodynamics has been used by various authors to study the
statistical properties of simple fluids subjected to nonequilibrium
constraints, such as
temperature gradient \cite{Others1,Others2,Others3} or shear
\cite{Dufty,Machta} (for a
review, see ref. \cite{schmitz98}).   Recent light scattering results,
obtained for systems
under temperature gradient, have shown quantitative agreement with
theoretical predictions
\cite{Sengers}.  Quantitative agreements have also been demonstrated with
results based on
particle simulations, both for systems under temperature gradient
\cite{Malek,BoonT} and shear
\cite{alexshear}. A more important issue is obviously the role of
fluctuations in the
onset of hydrodynamical instabilities, such as the convective instability
arising in the
B\'enard problem
\cite{benard}.

The macroscopic studies of sub-sonic
hydrodynamical instabilities are usually based on the incompressibility
assumption.
However, as first pointed out by Zaitsev and Shliomis \cite{Shliomis}, this
assumption is
basically inconsistent with the very foundation of the fluctuating
hydrodynamics formalism
since it imposes fictitious correlations between the velocity components of
the fluids.  On
the other hand, even in the absence of noise, the mathematical analysis of
convective
instabilities arising in compressible fluids proves to be quite involved
\cite{Cohen,CohenVB}. One
way to overcome this difficulty is to look for idealized models which, in
spite of their
extreme simplicity, can nevertheless lead to hydrodynamical instabilities
analog to those
observed in real systems. Our main purpose in this article is to study the statistical
properties of one such a model proposed some fifty years ago by Kolmogorov
\cite{obukhov}. As we will
show, the periodic boundary conditions associated to this model allow
detailed analysis of
the fluctuation spectrum from near equilibrium up to the vicinity of the
first instability
leading to convective rolls.

The Kolmogorov flow will be presented in the next section where some well
known aspects of its
macroscopic behavior are reviewed.  The statistical properties of the model
will be
discussed in section III.  We will show that the dynamic structure factor
of the fluid is
practically not affected by nonequilibrium constraints. This is not the
case for the velocity
correlation functions which become long-ranged as soon as the system is
driven out of
equilibrium.  Their amplitude is shown to diverges as one approaches to the
convective
instability threshold. Conclusion{\bf  s} and perspectives will be
presented in section IV.

\section{Kolmogorov flow}
\label{sec:II}

Consider an isothermal flow in a rectangular box $L_x \times L_y$ oriented
along the main
axes, that is
$\left\{ 0 \le x < L_x  ,  0 \le y < L_y \right\}$.  Periodic boundary
conditions are assumed
in both directions and the flow is maintained through an external force
field of the form
\begin{equation}
{\bf   F}_{ext}=F_0\,\sin{(2\,\pi\,n\,y/L_y)}\,{\bf   1}_x
\end{equation}
where ${\bf   1}_x$ is the unit vector in the $x$ direction. This model
represents the so-called  "Kolmogorov flow" and it belongs to the wider class of
two-dimensional negative effective eddy viscosity flows \cite{Gama}. It is
entirely
characterized through the strength of the force field $F_0$, the parameter
$n$, which
controls the wave number of the forcing, and the aspect ratio
$a_r$, defined as
\begin{equation}
a_r = L_x/L_y
\end{equation} In the following, we will mainly concentrate on the case
$n\,=\,1$.

The hydrodynamic equations for this model read:
\begin{equation}
\frac{\partial \,\rho}{\partial\,t} \, = \, -\grd\,\cdot\,(\rho\,{\bf   v})
\label{continuity}
\end{equation}
\begin{equation}
\rho\,\frac{\partial\,{\bf   v}}{\partial\,t} \, = \, -\rho\,({\bf
v}\cdot\grd)\,{\bf
v}\,-\,\grd\,P \,- \, \grd \cdot \bsigma \,+\, {\bf   F}_{ext}
\label{momentum}
\end{equation} where $\rho$ is the mass density, $P$ the hydrostatic
pressure and
$\bsigma$ the {\it two dimensional} stress tensor:
\begin{equation}
\sigma_{i, j} \, = \, - \, \eta \, \left(
\frac{\partial\,v_i}{\partial\,x_j} + \,
\frac{\partial\,v_j}{\partial\,x_i} - \, \delta_{i, j} \, \grd \cdot {\bf
v} \right) \, - \,
\zeta \, \delta_{i, j} \, \grd \cdot {\bf   v}
\label{stress}
\end{equation} For simplicity, we shall assume that the shear and bulk
viscosity coefficients,
$\eta$ and
$\zeta$, are {\it state independent}, i.e. they are constant.  It can then
be easily checked
that at the stationary state the pressure and the density are uniform in
space ($P_{st} =
P_0$, $\rho_{st} = \rho_0$) whereas the velocity profile is given by:
\begin{mathletters}
\begin{eqnarray}
{\bf   v}_{st}  =  & u_0  &\,\sin{(2\,\pi\,y/L_y)}\,{\bf   1}_x \\
 & u_0  &\, = \,\frac{F_0\,L_y^2}{4\,\pi^2\,\eta}
\end{eqnarray}
\end{mathletters} For small enough
$F_0$, this stationary flow is stable.  As we increase $F_0$, however, the
flow  may
become unstable giving rise to rotating convective patterns. Other
instabilities of
increasing complexity may occur for larger values of $F_0$, culminating in
a turbulent
- like behavior \cite{Lorenz,She,automata}. In this paper we shall
limit ourselves to the analysis of the system before its first instability.

We still have to supply the momentum conservation equation,
eq.(\ref{momentum}),  with an
equation of state relating the pressure to the density (recall that the
system is
isothermal).  In this section, we follow the traditional macroscopic
analysis by assuming
that the flow is incompressible, i.e.
\begin{equation}
\grd \, \cdot \, {\bf   v}\,=\, \frac{\partial{u}}{\partial x} \, + \,
\frac{\partial{v}}{\partial y} \, = 0
\label{incomp}
\end{equation} where $u$ and $v$ represent the $x$ and $y$ components of the velocity,
respectively, i.e.
${\bf   v} \equiv u {\bf   1}_x + v {\bf   1}_y$. Relation (\ref{incomp})
implies a
uniform density $\rho_0$ throughout the system for all times, if initially
so, as well as the
existence of a scalar function
$\psi (x, y)$, known as "stream function", defined by the relations:
\begin{equation}
u\,=\,\frac{\partial{\psi}}{\partial
y}\,,\,\,\,\,\,\,v\,=\,-\,\frac{\partial\psi}{\partial x}
\label{stream function}
\end{equation} Scaling lengths by $L_y$, velocities by $u_0 $  and time by
$L_y/u_0 $, the dimensionless equation for the stream function reads:
\begin{equation}
\frac{\partial(\nabla^2\,\psi)}{\partial t}\,=\,-\,\frac{\partial
\psi}{\partial y}\,\frac{\partial(\nabla^2\,\psi)}{\partial
x}\,+\,\frac{\partial\psi}{\partial x}\,\frac{\partial(\nabla^2\,\psi)}{\partial
y}\,+\,R^{-1}\,\nabla^2\,(\nabla^2\,\psi)\,+\,8\,\pi^3\,R^{-1}\,\cos{(2\,\pi
\,y) }
\label{incompsi}
\end{equation} where $R$ represents the Reynolds number,
\begin{equation}
R\, = \,\frac{u_0 \,L_y}{\nu}
\end{equation} and $\nu \equiv \eta / \rho$ is the kinematic viscosity. The
stationary
solution of (\ref{incompsi}) reads:
\begin{equation}
{\psi}_{st}\,=\,-\,\frac{1}{2\,\pi}\,\cos{(2\,\pi\,y)}
\end{equation} Setting $\psi = \psi_{st} + \delta \psi$, and linearizing
(\ref{incompsi})
around $\psi_{st}$, one gets
\begin{equation}
\frac{\partial (\nabla^2\,\delta\psi)}{\partial
t}\,=\,-\,\sin{(2\,\pi\,y)}\,\frac{\partial
(\nabla^2\,\delta\psi)}{\partial
x}\,-\,4\,{\pi}^2\,\sin{(2\,\pi\,y)}\,\frac{\partial\,\delta\psi}{\partial
x}\,+\,R^{-1}\,\nabla^2\,(\nabla^2\,\delta\psi)
\label{psilin}
\end{equation} Owing to periodic boundary conditions, $\delta
\psi(x, y, t)$ can be expanded in Fourier series:
\begin{eqnarray}
\delta\psi(x,\,y,\,t)&=&\sum_{k_x,\,k_y\,=\,-\,\infty}^{\infty}\exp{(-\,2\,\
pi\, i\,k_y\,y)}
\,\exp{(-\,2\,\pi\,i\,k_x\,x/a_r)}\,\delta\psi_{k_x,\, k_y}(t)\, , \nonumber\\
\delta\psi_{k_x,\, k_y}(t)&=&\int_{0}^{1}dy\,\exp{(2\,\pi\,i\,k_y\,y)}
\,\frac{1}{a_r}\int_{0}^{a_r}dx\,\exp{(2\,\pi\,i\,k_x\,x/a_r)}\,\delta\psi(x
,\,y ,\,t)
\nonumber\\
\end{eqnarray}  Equation (\ref{psilin}) can then be transformed to
\begin{eqnarray}
\frac{\partial \delta\psi_{k_x,\,k_y}}{\partial t} \, = & - &
4\,{\pi}^2\,R^{-1}\,({\tilde
{k}_x}^2\,+\,{k_y}^2)
\delta\psi_{k_x,\, k_y} \nonumber\\ & + & \pi\,\tilde{k}_x\,
\left[ \delta\psi_{k_x,\,k_y\,+\,1}\,-\,\delta\psi_{k_x,\,k_y\,-\,1}\right] \nonumber\\
 & + & 2\,\pi\,\frac{\tilde {k}_x\,k_y}{{\tilde {k}_x}^2\,+\,{k_y}^2}\,
\left[ \delta\psi_{k_x,\,k_y\,+\,1}\,+\,\delta\psi_{k_x,\,k_y\,-\,1} \right]
\label{fourlin}
\end{eqnarray}  where we have set
\begin{equation}
\tilde {k}_x = k_x/a_r
\end{equation}

In its general form, the analysis of this equation proves to be quite difficult
\cite{Sinai}.  On the other hand, if $\psi_{st}$ is  stable then, in the
long time limit,
the evolution of the system will be mainly governed by long wave length
modes. Accordingly, we
start our analysis by considering only the modes
$k_y\,=\,0\,,\pm 1$, i.e. we assume that $\delta\psi_{k_x,\,k_y}(t) \approx
0$ for
$|k_y|\,\geq 2$ \cite{green}. The
equation (\ref{fourlin}) then reduces to a
$3
\times 3$ matricial equation whose eigenvalues for  {\bf  $k_x \ne 0$} (the
case $k_x = 0$ is
trivial) are found to be :
\begin{eqnarray}
\lambda_1 \,&=&\,-\,2\,\pi^2\,R^{-1}\,(1\,+\,2\,{\tilde
{k}_x}^2)\,+\,\pi\,\sqrt{2\,{\tilde
{k}_x}^2\, (1-{\tilde {k}_x}^2) / (1 +{\tilde {k}_x}^2)
\,+\,4\,\pi^2\,R^{-2}} \,,\nonumber
\\
\lambda_2 \,&=&\,-\,2\,\pi^2\,R^{-1}\,(1\,+\,2\,{\tilde
{k}_x}^2)\,-\,\pi\,\sqrt{2\,{\tilde
{k}_x}^2\, (1-{\tilde {k}_x}^2) / (1 +{\tilde {k}_x}^2)
\,+\,4\,\pi^2\,R^{-2}} \,,\nonumber
\\
\lambda_3 \,&=&\,-\,4\,\pi^2\,R^{-1}\,(1\,+\,{\tilde {k}_x}^2)\,.
\label{vpropre}
\end{eqnarray}

It follows from (\ref{vpropre}) that  $\lambda_2$ and $\lambda_3$ are
always negative, whereas
there exists a critical value of the Reynolds number
\begin{equation}
R_c(k_x)\,=\,2\,\sqrt{2}\,\pi\,\frac{1\,+{\tilde
{k}_x}^2}{\sqrt{1\,-\,{\tilde {k}_x}^2}} \, \,  ; \, \, \, \, \, \, \, \, 0
\, < \, {\tilde {k}_x}^2 \, \le \, 1
\label{Rc}
\end{equation} for which $\lambda_1$ becomes equal to zero, thus indicating
the limit of the
stability of the corresponding mode \cite{reynoldscrit}.  Clearly
$R_c$ is an increasing function of $|k_x|$ so that the first modes to
become unstable
correspond to $|k_x| = 1$, provided the aspect ratio $a_r > 1$. As
$a_r\,\rightarrow\, 1$,
$R_c\,\rightarrow\,\infty$, indicating that no instability can develop for
perturbations of
the same spatial periodicity as the applied force (see ref. \cite{Mar}).  In the
following, we shall therefore concentrate on the case $a_r > 1$.

For $a_r = 2$, relation (\ref{Rc}) predicts a critical Reynolds number
$R_c = 5 \pi \sqrt{6}
/3 $ $\approx 12.8255$.  Analytical calculations can still be handled when
the modes
$k_y =
\pm 2,
\pm 3$ are taken into account as well, and lead to a critical Reynolds
number $R_c
\approx 12.8736$.  The discrepancy thus remains smaller than
$0.4 \%$.  Numerical evaluation of
$R_c$ performed with a total amount of $103$ modes shows no further
discrepancy.  We thus
conclude that one can rely reasonably well on a "3-modes" approximation
theory (that is
$\delta\psi_{k_x,\,k_y}(t) \approx 0$ for $|k_y|\,\ge 2$), as long as
$R$ is close to $R_c$. We shall use this approximation in the next section
to study the
statistical properties of the system near its first instability.

\section{Hydrodynamic fluctuations}
\label{sec:III}

To study the fluctuation{\bf  s} spectrum, we first linearize the
hydrodynamic equations
(\ref{continuity}, \ref{momentum}) around the stationary s{\bf  t}ate. Setting
$\rho \, = \,
\rho_0\,+\,\delta\rho$, $P \, = \, P_0\,+\,\delta P$ and
${\bf   v} \, = \,{\bf   v}_{st} \,+ \, \delta{\bf   v}$, and following
Landau and Lifshitz
\cite{Landau}, the fluctuating hydrodynamic equations read:
\begin{eqnarray}
\frac{\partial\,\delta\rho}{\partial t} & = & -\rho_0\, \left (
\frac{\partial\,\delta u}{\partial x}\,+\,\frac{\partial\,\delta v}{\partial y}
\right ) \,-\,u_0
\,\sin{(2\,\pi\,y/L_y)}\,\frac{\partial\,\delta\rho}{\partial x}\,, \\
\rho_0\,\frac{\partial\,\delta {\bf   v}}{\partial\,t} & = & -\rho_0\,({\bf
v}_{st}\cdot\grd)\,\delta {\bf   v}\,-\rho_0\,(\delta {\bf   v}
\cdot\grd)\,{\bf
v}_{st}\,-\,\grd\,\delta P \,- \, \grd \cdot \delta \bsigma
\end{eqnarray}
$\delta \bsigma$ is the two dimensional fluctuating stress tensor:
\begin{equation}
\delta  \sigma_{i, j} \, = \, - \, \eta \, \left( \frac{\partial\,\delta
v_i}{\partial\,x_j}
+ \,
\frac{\partial\,\delta v_j}{\partial\,x_i} - \, \delta_{i, j} \, \grd \cdot
\delta {\bf   v} \right) \, - \, \zeta \, \delta_{i, j} \, \grd \cdot
\delta {\bf   v} \, +
\, S_{i, j}
\label{flucstress}
\end{equation} where ${\bf   S}$ is a random tensor whose elements $\{
S_{i, j} \}$ are
Gaussian white noises with zero mean and covariances given by
\begin{eqnarray}
< S_{i, j}({\bf   r}, t) \, S_{k, \ell}({\bf   r}', t') > & = & 2 k_B T_0 \,
\delta (t - t')  \, \delta ({\bf   r} -{\bf   r}') \left [ \eta (\delta_{i,
k}^{Kr}
\delta_{j, \ell}^{Kr} +
\delta_{i,\ell}^{Kr} \delta_{j, k}^{Kr} ) \, + \, (\zeta - \eta) \delta_{i,
j}^{Kr}
\delta_{k, \ell}^{Kr} \right ]
\label{noisecor}
\end{eqnarray}

We still have to specify the equation of state.  Since the fluid is
isothermal, we simply set
\begin{equation}
\delta P\,=\,c_s^2\,\delta\rho\,,
\end{equation} where $c_s$ is the isothermal sound speed.  Scaling lengths
by $L_y$, time by
$L_y/c_s$, $\delta \rho$ by $\rho_0$ and $\delta {\bf   v}$ by the speed of
sound, the
dimensionless fluctuating equations in the Fourier space read:
\begin{eqnarray}
\frac{\partial\,\delta\rho_{k_x,\,k_y}(t)}{\partial t} & = & 2\,\pi\,i\,
\left( \tilde{k}_x\,\delta u_{k_x,\,k_y}\,+\,\,k_y\,\delta v_{k_x,\,k_y}
\right)
\, + \,
\epsilon\,R\,\pi\,\tilde{k}_x\,(\delta\rho_{k_x,\,k_y+1}\,-\,\delta\rho_{k_x
,\,k _y-1})\,,
\label{rofour}
\end{eqnarray}

\begin{eqnarray}
\frac{\partial \delta u_{k_x,\,k_y}(t)}{\partial t}   =  & - &
\pi\,\epsilon\,R(\delta
v_{k_x,\,k_y+1}\,+\,\delta v_{k_x,\,k_y-1}) \, + \,
\pi\,\epsilon\,R\,\tilde {k}_x\,(\delta u_{k_x,\,k_y+1}\,-\,\delta u_{k_x,\,k_y-1})
\nonumber\\
   & - &  \,4\,\pi^2\,\epsilon({\tilde {k}_x}^2\,+\,k_y^2)\,\delta
u_{k_x,\,k_y} \, -
\,4\,\pi^2\,\alpha\,\epsilon\,\tilde {k}_x (\tilde {k}_x\,\delta
u_{k_x,\,k_y}\,+\,
k_y\,\delta v_{k_x,\,k_y}) \nonumber\\
   & + & 2\,\pi\,i\,\tilde{k}_x\,\delta\rho_{k_x,\,k_y}\, + F_{k_x,\,k_y}(t)\,,
\end{eqnarray}

\begin{eqnarray}
\frac{\partial \delta v_{k_x,\,k_y}(t)}{\partial t}  = & &
\pi\,\epsilon\,R\,\tilde {k}_x\,(\delta v_{k_x,\,k_y+1}\,-\,\delta
v_{k_x,\,k_y-1})
\,-\,4\,\pi^2\,\epsilon({\tilde {k}_x}^2\,+\,k_y^2)\,\delta v_{k_x,\,k_y}
\nonumber \\
 & - & 4\,\pi^2\,\alpha\,\epsilon\, k_y\,(\tilde {k}_x\, \delta
u_{k_x,\,k_y}\,+\,k_y\,\delta
v_{k_x,\,k_y})\, + 2\,\pi\,i\,k_y\,\delta\rho_{k_x,\,k_y} \, + \,
G_{k_x,\,k_y}(t).
\label{complin}
\end{eqnarray} where $\alpha = \zeta/\eta$ and
\begin{equation}
\epsilon\,=\,\frac{\nu}{c_s\,L_y}\,.
\end{equation} The functions $F_{k_x,\,k_y}$ and $G_{k_x,\,k_y}$ are
Fourier components of
the noise terms ; their covariances follow directly from eqs. (\ref{flucstress},
\ref{noisecor}):
\begin{eqnarray}
< F_{k_x,\,k_y}(t)\,F_{k'_x,\,k'_y}(t') > & = & \epsilon \,
A\,[(\alpha\,+\,1)\,
\tilde{k}_x^2\,+\,k_y^2]\, \delta_{{\bf   k}+{\bf   k}',0}^{Kr}\,\delta
(t\,-\,t')\,,\nonumber\\ < F_{k_x,\,k_y}(t)\,G_{k'_x,\,k'_y}(t') > & = &
\epsilon \,
A\,\alpha\,
\tilde{k}_x\,k_y \, \delta_{{\bf   k}+{\bf   k}',0}^{Kr}\,\delta
(t\,-\,t')\,,\nonumber\\ <
G_{k_x,\,k_y}(t)\,G_{k'_x,\,k'_y}(t') > & = & \epsilon \, A\,[{\tilde
{k}_x}^2\,+\,(\alpha\,+\,1)\,k_y^2]\, \delta_{{\bf   k}+{\bf
k}',0}^{Kr}\,\delta
(t\,-\,t')\,,
\end{eqnarray} where ${\bf   k} \equiv (k_x/a_r , \, k_y)$ and
\begin{equation}
A\,=\, 8\,\pi ^2 \, \frac{k_B\,T_0}{M \,{c_s}^2} \, ,
\end{equation}
$M \, = \, a_r\,\rho _0 \,{L_y}^2$ being the total mass of  the system. \\

The analysis of the above Langevin equations can be simplified somewhat by
noticing that the
quantity $\epsilon$ must remain small if one wishes to remain within the
limit of validity of
the hydrodynamic regime \cite{Alder}. Furthermore, as already mentioned in
the introduction,
in this article we limit ourselves to strictly sub-sonic flows, so that
$\epsilon R = u_0 / c_s << 1$.  We thus have at our disposal a natural
small parameter which,
however, have to be used with care since the solution of the Langevin
equations (\ref{rofour}
- \ref{complin}) proves to be singular in the limit $\epsilon
\rightarrow 0$. Moreover, it turns out that the behavior of the system is
qualitatively
independent of the value of the bulk viscosity coefficient so that, to
avoid cumbersome
notations, we simply set $\alpha = 0$ (recall that $\alpha = \zeta/\eta$).
In any case, the
calculations remain lengthy and tedious, so that here we concentrate mainly
on the final
results, giving only a brief sketch of the intermediate steps.

We pay particular attention to two quantities.  First, the so-called
scattering function,
defined as the  space-time Fourier transform of the density
auto-correlation function:
\begin{equation}
S_{\bf   k}(\omega) \, = \, \int_{- \infty}^{+ \infty} dt\,\exp{(i \, \omega
t)}
\,<\delta \rho_{{\bf   k} }(t) \, \delta \rho_{-{\bf   k} }(0)>
\end{equation}
\begin{equation}
<\delta \rho_{{\bf   k} }(t) \, \delta \rho_{-{\bf   k} }(0)> \, = \,
\frac{1}{S^2} \, \int \! \! \! \int d{\bf   r} \,d{\bf   r}' \,\exp{\left\{
2\,\pi\,i
\,{\bf   k}
\cdot  ({\bf   r} - {\bf   r}') \right\} }
\, <\delta \rho( {\bf   r}, \, t) \, \delta \rho( {\bf   r}', \, 0)>
\label{FTro}
\end{equation} where the integrals extend over the surface $S =  a_r \times
1$ of the system.
Next, the space-time Fourier transform of the velocity auto-correlation
function, defined in
a similar fashion:
\begin{equation}
W_{\bf   k}(\omega) \, = \, \int_{- \infty}^{+ \infty} dt\,\exp{(i \, \omega
t)}
\,<\delta {\bf   v}_{{\bf   k} }(t) \, \cdot \, \delta {\bf   v}_{-{\bf
k} }(0)>
\end{equation} as well as its static (equal time) counterpart:
\begin{equation}
<\delta {\bf   v}_{{\bf   k} } \, \cdot \, \delta {\bf   v}_{-{\bf   k} }>
\, = \,
\frac{1}{2
\, \pi }\, \int_{-
\infty}^{+ \infty} d \omega \,
\,W_{\bf   k}(\omega)
\end{equation}

We found that, to dominant order in $\epsilon$, $S_{\bf   k}(\omega)$ is
not affected by the
nonequilibrium constraints, i.e.
\begin{equation}
S_{\bf   k}(\omega) \, = \, {\frac { 32 \, \epsilon\,A \, {k}^{4} \, {\pi
}^{4} }{\left(
 {\omega}^{2} - 4\,{k}^{2}\,{\pi }^{2}\right )^{2}+16\,{\epsilon}^{2} \,
{\omega}^{2} \, {k}^{
4}\,{\pi }^{4} }\, } \, \left[ 1 \, + \, O(\epsilon^2 R^2) \right]
\label{skw}
\end{equation} where $k^2 \equiv (\tilde {k}_x^2 + k_y^2)$.  We note that
the scattering
function exhibits only sound mode peaks (Brillouin lines). The absence of a
purely
dissipative mode around
$\omega \approx 0$ (Rayleigh line) is directly related to the fact that the
Kolmogorov flow
is "strictly" isothermal, i.e. there are no temperature (or entropy)
fluctuations.  On
the other hand, the velocity auto-correlation function do exhibit a  purely
dissipative
viscous regime around $\omega \approx 0$, together with a sound regime
located around $\omega
\approx
\pm \, 2 \pi |k|$.  Here again we found that, to dominant order in
$\epsilon$, the sound regime is not affected by the nonequilibrium
constraints and behaves
very much like the scattering function, eq. (\ref{skw}).  One thus arrives
at the conclusion
that the nonequilibrium constraints affect mainly the behavior of the fluid
near the origin
$\omega
\approx 0$ (the viscous regime).

We first consider near equilibrium situations, limiting ourselves to
relatively small values
of the Reynolds number $R$. In this case the Langevin eqs.
(\ref{rofour}~-~\ref{complin}) can
be solved perturbatively by expanding the hydrodynamic variables around the
equilibrium,
\begin{mathletters}
\begin{eqnarray}
\delta \rho = & \delta \rho_{eq} + \mu \, \delta \rho_1 + \mu^2 \, \delta
\rho_2 + \dots \\
\delta {\bf   v} = & \delta {\bf   v}_{eq} + \mu \, \delta {\bf   v}_1 +
\mu^2 \,
\delta {\bf   v}_2 +
\dots
\end{eqnarray}
\end{mathletters} where the subscript "$eq$" denotes equilibrium quantities
and the parameter
$\mu$ is defined as $\mu \equiv R/2\pi$.  After some tedious algebra, one
gets for the
static correlation function:
\begin{eqnarray}
<\delta {\bf   v}_{{\bf   k} }  \cdot  \delta {\bf   v}_{-{\bf   k} }> \, -
\, <\delta {\bf   v}_{{\bf   k} } \cdot  \delta {\bf   v}_{-{\bf   k}
}>_{eq} \,  =  \,
\nonumber
\end{eqnarray}
\begin{equation}
\left (R/2\pi\right)^2 \,\frac { A\left (10+2\, \tilde {k}_x^{6}+5\, \tilde
{k}_x^{4}+\tilde
{k}_x^{2}\right )}{2
\left(\tilde {k}_x^{2}+4\right )\left (2
\,\tilde {k}_x^{2}+5\right )\left (2\,\tilde {k}_x^{2}+1\right )\left (\tilde
{k}_x^{2}+1\right )^{2} }\left[ 1 \, + \, O(\epsilon^2,\, (R/2\pi)^2) \right]
\label{vdotv}
\end{equation} where $<\delta {\bf   v}_{{\bf   k} } \cdot  \delta {\bf
v}_{-{\bf   k}
}>_{eq}
\, = \, 2 A$ is the equilibrium contribution and, to simplify the
presentation, we have
considered the case $k_y = 1$.

To check the validity of this result, we have solved numerically the
Langevin eqs.
(\ref{rofour}~-~\ref{complin}).  The traditional procedure consists of
simulating the
corresponding stochastic processes and using the hydrodynamical sample
paths (time series) so
obtained to construct the various correlation functions. Unavoidable for
nonlinear problems,
this procedure is quite simple to set up but requires very long runs in
order to get reliable
statistics. Alternatively, one can solve directly the equations governing
the evolution of
the correlation functions which can be obtained easily from the underlying
Langevin equations
\cite{alexjstatphys}.  This latter procedure is accurate (no
need for statistics) and quite fast but it is, of course, limited to linear
problems.
We have used both techniques,  the former to simulate the full nonlinear
hydrodynamic
equations (\ref{continuity},
\ref{momentum}) with noisy stress tensor, and the latter to study the
statistical properties
of the (linear) Langevin equations (\ref{rofour}~-~\ref{complin}), limiting
ourselves to the
first 41 $k_y$ modes (that is
$\delta\rho_{k_x,\,k_y}(t) = \delta{\bf   v}_{k_x,\,k_y}(t) \approx 0$ for
$|k_y|\,\ge 21$).

In figure 1 we have presented the static velocity auto-correlation
function, as given by the
relation (\ref{vdotv}), together with the corresponding numerical solution.
As it can
be seen, quantitative agreement is demonstrated for $R \le 4$ but
discrepancies gradually
appear as we consider larger values of the Reynolds number.  This is to be
expected
since the validity of the relation (\ref{vdotv}) can only be guaranteed for
"small"
values of the Reynolds number.

Before discussing the behavior of the system for larger values of $R$ ($<
R_c$) it is
instructive to study the properties of the static correlation function in
real space,
$<\delta {\bf   v}({\bf   r}) \, \cdot \, \delta {\bf   v}(0)>$.  This can
be obtained by
summing the product  $<\delta {\bf   v}_{{\bf   k} }  \cdot
\delta {\bf   v}_{-{\bf   k} }>$ $\times$  $\exp [ 2 \pi i (x \, \tilde
{k}_x + y \, k_y) ]$
over ($k_x, \, k_y$).  Analytical calculations, however, prove to be
extremely difficult to
handle for the general case.  We therefore limit ourselves to a special
case where only one
of the wavenumber is summed over, the other being held fixed. Specifically,
we simply set
$k_y = 0$ to obtain:
\begin{equation}
<\delta {\bf   v}_{k_x}  \cdot  \delta {\bf   v}_{-{k_x}}> \, -  \, 2 A \,  =
\, \left (R/2\pi\right)^2
\,\frac {A/2}{\left (\tilde {k}_x^{2}+1\right )\left (1+2\,\tilde
{k}_x^{2}\right )}\left[ 1
\, + \, O(\epsilon^2,\, (R/2\pi)^2)
\right]
\label{ky0}
\end{equation} Note that setting $k_y = 0$ is equivalent to taking the
spatial average over
the
$y$ direction (cfr. eq. (\ref{FTro})), so that relation (\ref{ky0}) holds
only for
$k_x \ne 0$.  In fact, $\delta {\bf   v}_{0, \, 0}(t) \equiv 0$ since the
linear momentum of
the center of mass is a conserved quantity. With this restriction in mind,
the summation can
be performed in a straightforward manner to give
\begin{eqnarray}
 < \delta {\bf v}(x) \cdot \delta {\bf v}(0) > & - & 2\,A\,[\delta (x)\,-\,a_r]
  \, = \, \nonumber\\
 \frac{A\,R^2\,a_r}{8\,\pi}\, \left[\sqrt{2}\,\frac{\cosh[\sqrt{2}
 \,\pi\,(x - a_r/2)]}{\sinh(\pi\,a_r/\sqrt{2})} \right .
 & - & \left .\frac{\cosh[2\,\pi\,(x -
a_r/2)]}{\sinh(\pi\,a_r)}\,-\,\frac{1}{\pi\,a_r}
\right]
\label{corx}
\end{eqnarray} where the second term on the left hand side is the
equilibrium contribution
\cite{malek87}.  Note the presence of a constant term in both equilibrium
and nonequilibrium
(right hand side) part which ensures the conservation of the linear momentum.

The nonequilibrium contribution to $<\delta {\bf   v}(x) \cdot \delta{\bf
v}(0) >$ exhibits
long-range correlations since the correlation length is clearly of the order of
system's size.  This is shown in figure (2) for $R=3$ where quantitative
agreement with
numerical results is observed.  The existence of long-range correlations is
generic for
fluids under shear constraints and have been predicted by  several authors
\cite{Dufty,malek87}, and confirmed by both microscopic \cite{alexshear}
and lattice-gas
automata simulations
\cite{BoonU}.  On the other hand, experimental evidence has been so far
reported only for
fluids under a temperature gradient, where quantitative agreement with
fluctuating
hydrodynamics was demonstrated
\cite{Sengers}.

Let us now consider far from equilibrium situations. As pointed out in the
last section, for
$R$ close to
$R_c$ one can reasonably well rely on the "3-modes" approximation theory
(that is
$\delta\rho_{k_x,\,k_y}(t) = \delta{\bf   v}_{k_x,\,k_y}(t) \approx 0$ for
$|k_y|\,\ge 2$).  As a consequence, equations (\ref{rofour} -
\ref{complin}) reduce to a
system of nine coupled linear Langevin equations for each fixed $k_x$
which, for consistency,
must be limited to $|k_x| \le a_r$.  The calculations can nevertheless be
done, leading to
the following expression for the static velocity auto-correlation function:
\begin{equation}
<\delta {\bf   v}_{{\bf   k} }  \cdot  \delta {\bf   v}_{-{\bf   k} }> \, -
\, 2 A
\,  =  \,\frac {A \, R^2}{2
\left(R_ c^{2}-R^2 \right ) \left (1+2
\,\tilde {k}_x^{2}\right )}\left[ 1 \, + \, O(\epsilon^2 \, R^2) \right]
\label{corcrit}
\end{equation} where the second term on the left hand side is the
equilibrium contribution,
$R_c(k_x)$ is given by eq. (\ref{Rc}) and  $k_y = 1$.  The nonequilibrium
part diverges as $R
\rightarrow R_c(k_x=a_r)$, but then, of course, the linearized Langevin
equations ceases to
be valid.  In figure 3 we have represented the result (\ref{corcrit}) for
increasing values
of $R$, together with the numerical solution of the linear Langevin
equations (\ref{rofour} -
\ref{complin}) as well as the results obtained by simulation of the full
nonlinear
hydrodynamic equations (\ref{continuity}, \ref{momentum}) with noisy stress
tensor.
Quantitative agreement is observed for values of $R$ up to $12$, but
significant discrepancies
start to show up as $R \rightarrow R_c (\approx 12.87)$ where the
linearized theory leads to
diverging correlation function (cfr. eq. (\ref{corcrit})).  This is not the
case for the
correlation function based on the full nonlinear equations which seems to
exhibit a maximum
around $R_c$.  It should however be noted that, due to slowing down of the
relaxation of
the "critical" Fourier modes, statistical errors are quite important for
$R$ close to
$R_c$ (about 15\% for the last four data), so that no definitive conclusion
can be drawn at
this stage.  In any case, the analysis of the statistical properties of the
nonlinear regime
is beyond the scope of the present work.

\section{Validity of the incompressibility assumption}

The macroscopic studies of sub-sonic hydrodynamical instabilities are based on
the incompressibility assumption which is fundamentally inconsistent with
the very
foundation of the fluctuating hydrodynamics formalism \cite{Shliomis}. For
instance, it
is easy to show that at equilibrium ($R=0$), one has:
\begin{equation}
\frac{U_{\bf   k}(\omega)_{eq}} { V_{\bf   k}(\omega)_{eq} } \, = \,\frac
{\left(
\omega^2 - 4\,\pi^2 k_y^2 \right)^2+16 \, \pi^4
\, \tilde {k}_x^2 k_y^2}{\left( \omega^2 - 4\,\pi^2 \tilde {k}_x^2
\right)^2+16 \, \pi^4
\, \tilde {k}_x^2 k_y^2} \, \left[ 1 \, + \, O(\epsilon^2) \right]
\label{u/v_equil}
\end{equation} where $U_{\bf   k}(\omega)$ and $V_{\bf   k}(\omega)$ are
the space-time
Fourier transforms of
$~\!\!\!<\delta  u({\bf   r}, t)  \delta  u({\bf   r}', 0)~\!\!\!>$ and
$<~\!\!\!\delta  v({\bf   r}, t)  \delta v({\bf   r}', 0)~\!\!\!>$
respectively.  On the
other hand, the incompressibility assumption, eq. (\ref{incomp}), implies
\begin{equation}
U_{\bf   k}(\omega)_{inc} \, /  \, V_{\bf   k}(\omega)_{inc} \, = \, k_y^2
\, /
\, \tilde {k}_x^2
\label{u/v_incomp}
\end{equation} where the subscript "{\it inc}" refers to incompressible
fluids.   Except near
the origin $\omega
\approx 0$ (long time limit), this result is clearly in contradiction with
the correct
equilibrium form, eq. (\ref{u/v_equil}).  In particular, the equilibrium static
auto-correlations are independent of the wave vector,
\begin{equation}
<\delta u_{{\bf   k} }\,  \delta u_{-{\bf   k} }>_{eq} \, = \, <\delta
v_{{\bf   k} }\,  \delta v_{-{\bf   k} }>_{eq} \, = \, A
\label{statequil}
\end{equation} whereas relation (\ref{u/v_incomp}) leads to  $<\delta
u_{{\bf   k} }\,  \delta
u_{-{\bf   k} }>_{inc}/ <\delta v_{{\bf   k} }\, \delta v_{-{\bf   k}
}>_{inc}\, = \, k_y^2
\, / \, \tilde {k}_x^2 $.

The situation is somewhat different for the nonequilibrium case. As we have
shown in the
previous section, to dominant order in
$\epsilon$ both the scattering function and  the sound regime of the velocity
auto-correlation function assume their equilibrium form, regardless of the
value of the
Reynolds number.  The nonequilibrium constraints thus affect mainly the
behavior of the fluid
near the origin
$\omega \approx 0$ (the viscous regime).  This result has an interesting
consequence.  It
suggests that, as far as the nonequilibrium properties of the fluid are
concerned, one may
rely on the incompressibility assumption, eq. (\ref{incomp}), since the
compressibility of
the fluid affects mainly the sound modes which are well separated from the purely viscous
modes (recall that $\epsilon$ is small).  Analytical calculations confirm
the above arguments
and lead to the following relation:
\begin{equation}
\frac{\, U_{\bf   k}(\omega) \, - \, U_{\bf   k}(\omega)_{eq}} {V_{\bf
k}(\omega) \, - \,
V_{\bf   k}(\omega)_{eq} } \, = \, \frac{ k_y^2}{\tilde {k}_x^2} \,
\left[ 1
\, + \, O(\epsilon^2\,R^2) \right]
\label{u/v}
\end{equation} where both the numerator and the denominator in the left
hand side prove to
assume a Lorentzian shape, sharply peaked around the origin (the width is
of the order of
$\epsilon$), since the sound regime cancels out. Nevertheless, because of
the presence of the
equilibrium contributions, the relation (\ref{u/v}) is still in
contradiction with the
incompressibility condition, eq. (\ref{u/v_incomp}).  There exist, however,
two different
situations where this objection can be ruled out.

First, near the origin ($\omega \approx 0$) where the fluid satisfies the
incompressibility
condition already at equilibrium, i.e.
$U_{\bf   k}(\omega)_{eq}\,{\tilde {k}_x^2}
\approx V_{\bf   k}(\omega)_{eq} \, k_y^2$.  Obviously, this situation
concerns only the long
time behavior of the fluid. For instance, the static correlation functions obey
\begin{equation}
\frac{<\delta u_{{\bf   k} }\,  \delta u_{-{\bf   k} }> \, - \,A} {<\delta
v_{{\bf   k} }\,
\delta v_{-{\bf  k} }>
\, - \, A} \, = \, \frac{ k_y^2}{\tilde {k}_x^2} \,
\left[ 1
\, + \, O(\epsilon^2\,R^2) \right]
\label{u/vstat}
\end{equation} which contradicts the incompressibility condition, eq.
(\ref{u/v_incomp}).

A more interesting situation concerns the behavior of the fluid near the
instability where it
can be shown that in the limit $R \rightarrow R_c$ both the static and
dynamic velocity
correlation functions behave as
$O((R_c^2
\, -
\, R^2)^{-1})$.  In other words, for $R$ "close" enough to
$R_c$,  equilibrium contributions become negligible so that the fluid
behaves basically in an
incompressible way.

It should however be realized that this appealing conclusion is based on
the linearized
Langevin equations (\ref{rofour} - \ref{complin}) which are not valid near
the convective
instability. The study of the statistical properties of the system in the
critical regime
requires a nonlinear analysis of the fluctuating equations which is beyond
the scope of the
present work and will be reported elsewhere. Instead, we resort here to
numerical analysis
only.  More precisely, we have simulated the full nonlinear fluctuating
hydrodynamic
equations to obtain the ratio of the
$x$ and $y$ components of the static velocity auto-correlation function for
several values of
the Reynolds number.  The results are depicted in figure 4 for {\bf $k_x =
k_y = 1,
 \, a_r = 2$}, so that the expected value of this ratio for an
incompressible fluid is
$4$.  This is precisely what we observe, but only for values of $R \ge
12.8$ (recall that
$R_c \approx 12.87$), a domain which is beyond the validity of linearized
hydrodynamic
equations (cfr. previous section).

\section{Concluding remarks}

Our main purpose in this article was the study of the statistical
properties of the
linearized Kolmogorov flow. The simplicity of this model allows a detailed
analysis of the
fluctuations spectrum from near equilibrium up to the vicinity of the first
instability
leading to convective rolls.  For this latter case, the analytical
calculations were based on
a "3-modes" approximation theory, which consists in retaining only the
Fourier modes with
wavenumber $|k_y|\,\le 1$, while for the former case we have set up a
perturbation scheme
around the equilibrium.  Extensive numerical calculations allow us to
emphasize clearly the
limit of validity of both regimes. In particular, we have shown that the
"3-modes"
approximation theory holds already for $R \ge 8$ and leads to a diverging
velocity
auto-correlation function as $R \rightarrow R_c$. On the other hand, the
simulation of the
full nonlinear fluctuating hydrodynamic equations indicates that the
validity of linearized
hydrodynamic equations can be guaranteed for Reynolds number as high as 12 ($R_c
\approx 12.87$).

Another interesting result concerns the validity of the incompressibility
assumption which
greatly simplifies the mathematical analysis of the problem. The
compressibility of a fluid
affects mostly fast sound modes whereas the dynamic of the system near an
instability is
expected to be governed mainly by dissipative slow modes.  This intuitive
argument has been
used by many authors who have considered fluctuating incompressible
hydrodynamic equations,
or even directly the corresponding normal form amplitude equations to which
they added random
noise terms \cite{noisybenard}.  In these approaches, the characteristics
of the noise terms
cannot be related to equilibrium statistical properties of the fluid and
thus remain
arbitrary.  A more satisfactory approach would be to start with the
compressible fluctuating
hydrodynamic equations. Such a procedure, however, proves to be extremely
difficult
mainly because of the boundary value problem.  To our knowledge, the only
attempt in this
direction has been made by Schmitz and Cohen for the case of B\'enard
instability
\cite{Cohen}.  Concentrating on the behavior of a small layer in the bulk,
these authors have
succeeded in deriving the linearized fluctuating equations close to the
convective
instability.

Here again, the relative simplicity of the Kolmogorov flow allows some
further progress in
this important issue. In this respect, we have shown that in the long time
limit the flow
behaves as an incompressible fluid, regardless of the value of the Reynolds
number.  This,
however, proves to be not the case for the short time behavior.  In
particular, the
incompressibility assumption leads in general to  a wrong  form of the static
correlation functions.  The only exception is near the convective
instability, where we have
shown that the incompressibility assumption remains valid.

The problem with this conclusion is that the linearized fluctuating
hydrodynamic equations,
on which this whole paper is based, are no more valid close to the
instability threshold.
Although extensive numerical simulations have basically confirmed our
predictions, a full
answer to this important problem requires nevertheless a nonlinear analysis
of the
fluctuating Kolmogorov flow.  Work on this direction is in progress.

\acknowledgments{We are very grateful to professors E. G. D. Cohen, G.
Nicolis, J. W. Turner
and C. Van den Broeck for helpful comments.  One of us (I. B.) acknowledges
successive support
from the EC TEMPUS program, the Center for Nonlinear Phenomena and Complex
Systems of the
Free University of Brussels, the GCIR of the French Community in Belgium,
and the IUAP, Prime
Minister's Office of the Belgian Government.}

\begin{figure}
\label{fig1}
\caption{ Fourier transform of the nonequilibrium part of the static
velocity auto-correlation
function, normalized by the corresponding equilibrium part, as a function
of the Reynolds
number. The solid curve represents the theoretical prediction, as given by
eq. (35), whereas
the diamonds correspond to results obtained by numerical simulation of the
linear Langevin
equations (23 - 25).  The parameters are : $\epsilon = 10^{-2}, A =
10^{-3}$ (defined by eq.
(28)), $k_y = 1, k_x = 1$  (i.e. $\tilde {k}_x = 1/2$).  The estimated
statistical errors are
less than $4 \%$.}
\end{figure}

\begin{figure}
\label{fig2}
\caption{ Nonequilibrium part of the static velocity auto-correlation
function, normalized by the parameter $A$ (defined by eq. (28)), as a
function of the spatial
coordinate $x$. The solid curve represents the theoretical prediction, as
given by eq. (37),
whereas the diamonds correspond to results obtained by numerical simulation
of the linear
Langevin equations (23 - 25). The Reynolds number is set to $R=3$ and the
other parameters
are as given in the caption of figure (1).}
\end{figure}

\begin{figure}
\label{fig3}
\caption{Fourier transform of the nonequilibrium part of the static
velocity auto-correlation
function, normalized by the corresponding equilibrium part, as a function
of the Reynolds
number. The solid curve represents the prediction based on the "3-modes"
approximation theory, eq. (38), whereas crosses and diamonds correspond to
numerical
results obtained respectively by the simulation of linear and nonlinear Langevin
equations, eqs. (3, 4).  The parameters are as given in the caption of
figure (1). The
estimated statistical errors are about $15 \%$ for the last four data points.}
\end{figure}

\begin{figure}
\label{fig4}
\caption{Ratio of $x$ and $y$ component of the velocity auto-correlation
function in Fourier
space, as a function of the Reynolds number.  The parameters are :
$\epsilon = 10^{-2}, A =
10^{-6}$ (defined by eq. (28)), $k_y = 1, k_x = 1$  (i.e. $\tilde {k}_x =
1/2$).  The dashed
line represents the expected ratio for an incompressible fluid, whereas the
diamonds
correspond to results obtained by the simulation of the full nonlinear
Langevin equations, eq.
(3, 4).  The estimated statistical errors do not exceed $8 \%$. }
\end{figure}

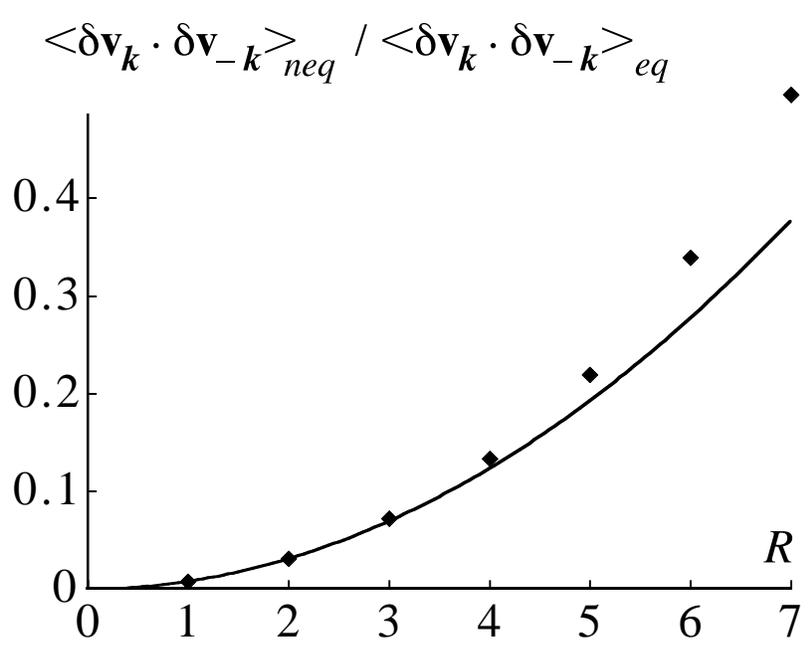
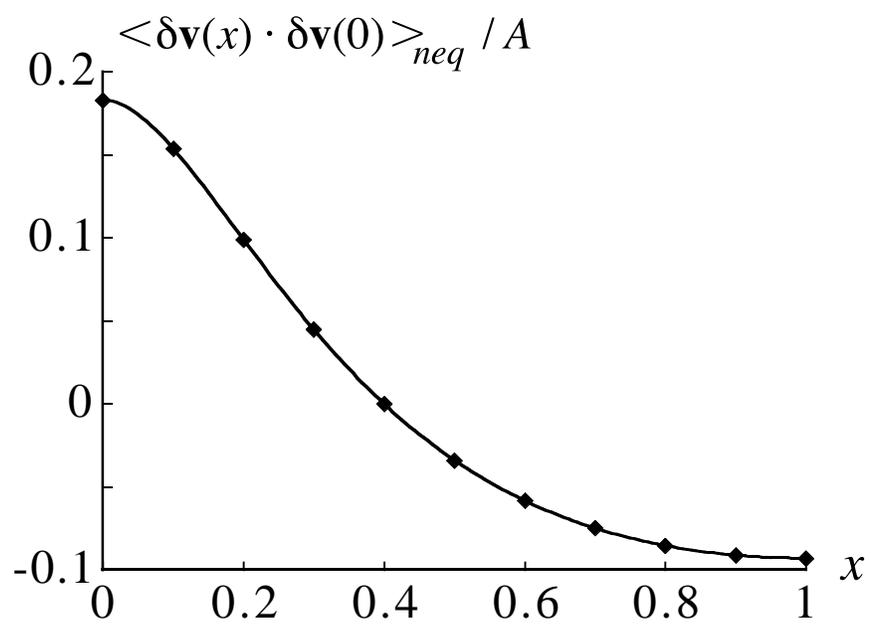
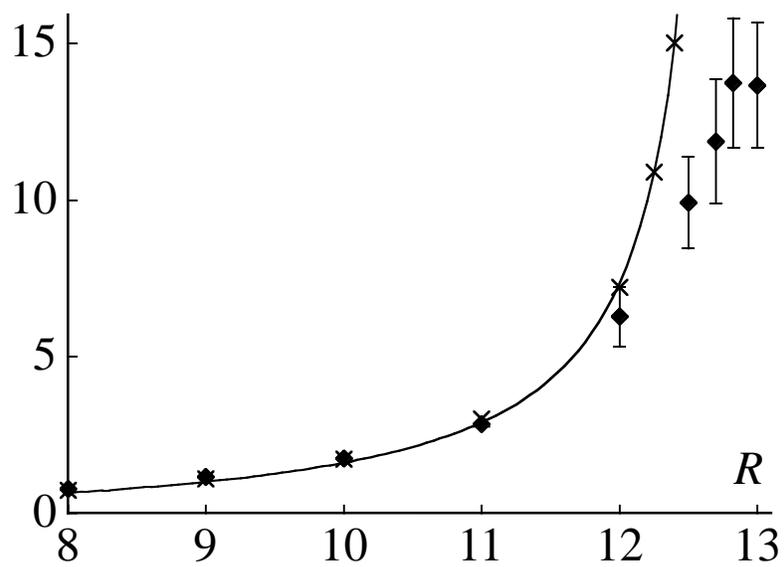
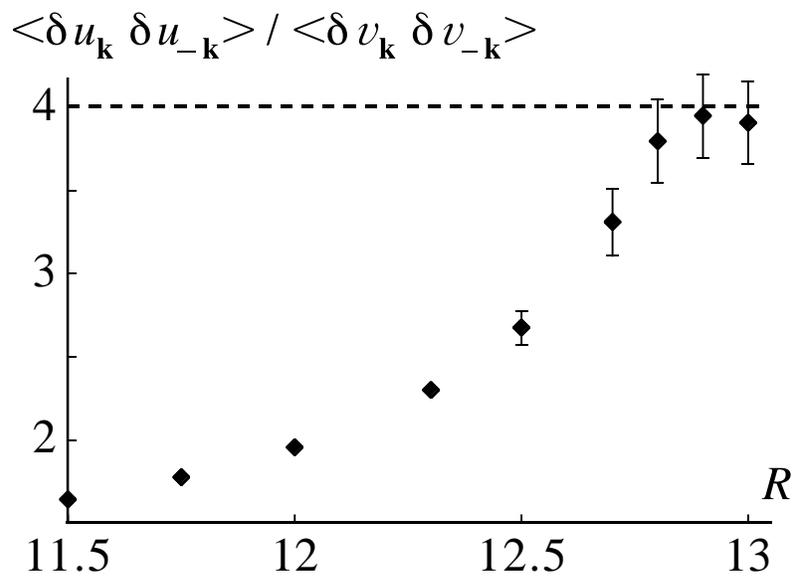
\end{document}